\documentclass[twocolumn]{jetpl}
\usepackage{mathtools}
\usepackage{bm}
\usepackage{graphicx} 
\usepackage{hyperref}
\usepackage[all]{hypcap}
\usepackage[sort&compress,numbers]{natbib}

\title{A method for measuring deviation from Lambert diffuse scattering law of ultracold neutrons on material walls}
\rtitle{A method for measuring deviation from Lambert diffuse scattering law of UCN on material walls}
\sodtitle{A method for measuring deviation from Lambert diffuse scattering law of UCN on material walls}

\author{P.\,D.~Grigoriev$^{abc}$\thanks{e-mail: grigorev@itp.ac.ru}, 
	V.\,D.~Kochev$^{bad}$, 
	V.\,A.~Tsyplukhin$^{b}$, 
	A.\,M.~Dyugaev$^{a}$,
	I.\,Ya..~Polishchuk$^{de}$}
\rauthor{P.\,D.\,Grigoriev, 
	V.\,D.\,Kochev, 
	V.\,A.\,Tsyplukhin, 
	A.\,M.\,Dyugaev,
	I.\,Ya.\,Polishchuk}
\sodauthor{Grigoriev, Kochev, Tsyplukhin, Dyugaev, Polishchuk}

\address{
	$^a$L.\,D.~Landau Institute for Theoretical Physics, 142432, Chernogolovka, Russia\\~\\
	$^b$Theoretical Physics and Quantum Technologies Department, National University of Science and Technology ''MISIS'', 119049, Moscow, Russia\\~\\
	$^c$National Research University Higher School of Economics, Moscow, 101000, Russia\\~\\
	$^d$NRC Kurchatov Institute, 123182, Moscow, Russia\\~\\
	$^e$Theoretical Physics Department, Moscow Institute For Physics and Technology, 141700, Dolgoprudnii, Russia}

\abstract{
	Modeling the motion of ultracold neutrons (UCNs) is crucial for assessing their losses, accurately measuring their lifetime, and describing other experiments. In material traps, it is necessary to account not only for specular but also for diffuse elastic reflection of UCNs from the trap walls. Typically, the Lambert cosine law is used to describe the angular distribution of diffusely scattered neutrons. However, this law lacks a rigorous theoretical derivation and is often violated. In our work, we propose an experiment to measure the deviation of the angular distribution of UCNs during diffuse scattering from the Lambert law. This deviation can be determined by the difference in the number of neutrons exiting through the central and end windows of a long narrow UCN trap. Monte Carlo simulations corresponding to a possible experiment have been performed, demonstrating a significant effect for different trap geometries.
}

\begin{document}
\maketitle

\textbf{1. Введение.}

Accurate measurements of neutron lifetime, 
$\tau_\text{n}\approx 15$ minutes, are essential for determining the weak interaction constants 
in nucleons and for various aspects of astrophysics and cosmology 
\cite{Dubbers/2011,Abele/2008,GONZALEZALONSO2019165,Universe2023}. 
The most precise measurements of neutron lifetime are based on experiments with ultracold neutrons (UCN) confined in material \cite{Serebrov2008PhysRevC.78.035505,Serebrov2018PhysRevC.97.055503} or magnetic traps \cite{Gonzalez2021,Ezhov2018}. The accuracy of these measurements is mainly limited by uncertainties in estimating UCN losses, as the measured trap lifetime 
$\tau$ is shorter by approximately $\sim 20$ seconds than the derived $\tau_\text{n}$ after accounting for these losses, even in the most precise experiments. Discrepancies between $\tau_\text{n}$ values obtained by these two methods significantly exceed their stated uncertainties \cite{Serebrov2018PhysRevC.97.055503,Gonzalez2021}. Moreover, a well-known unresolved discrepancy persists with neutron beam experiments \cite{BeamPhysRevC.71.055502,BeamPhysRevLett.111.222501,BeamReview2020}, potentially attributable to systematic errors in beam experiments \cite{Serebrov2021PhysRevD.103.074010}. Unaccounted losses in UCN material and magnetic traps, along with inaccuracies in estimating their rates, also remain likely \cite{Ezhov2023,grigoriev2024lossrateultracoldneutrons}.

A potential qualitative method to reduce UCN absorption involves coating trap walls with liquid $^{4}$He, the only material that does not absorb neutrons \cite{Golub1983,Bokun/1984,Alfimenkov/2009,Grigoriev2016Aug,GrigorievPRC2021,Grigoriev2021,GrigorievPRC2023}. However, this approach has several drawbacks \cite{GrigorievPRC2021,Grigoriev2021,GrigorievPRC2023}: (i) the low neutron potential barrier $V_{0}$, reducing UCN concentration, (ii) the necessity for low temperatures 
$T<0.5$~K to suppress UCN scattering by helium vapor, and (iii) significant UCN scattering by ripplons even at such low temperatures. Consequently, this study focuses on conventional UCN traps with absorbing walls.

Monte Carlo simulations are employed to enhance the accuracy of UCN loss rate estimates, incorporating gravity in material \cite{Fomin2023,Fomin2019,Ayres2018,Fomin2018,Fomin2017,Serebrov2013MC,Serebrov2018PhysRevC.97.055503} and magnetic \cite{Gonzalez2021,GONZALEZALONSO2019165} traps. These calculations address uncertainties in current approximation-based methods for determining UCN losses in material traps \cite{grigoriev2024lossrateultracoldneutrons}.

Monte Carlo modeling uses physical approximations for neutron-wall interactions. The probability of neutron absorption by the wall, 
$\mu (v_{\perp})$, depends on the incidence angle and is given by the formula \cite{Ignatovich/1990,Golub/1991,Ignatovich1996}:
\begin{equation}
	\mu (v_{\perp}) =
	\frac{2\eta \ v_{\perp}/v_{\lim} }{\sqrt{1-(v_{\perp}/v_{\lim})^{2}}},  \label{mu}
\end{equation}
where 
$\eta\sim 10^{-5}$ is the material-dependent loss coefficient, 
$v_{\perp}$ is the component of neutron velocity normal to the wall, and 
$v_{\lim}=\sqrt{2V_0/m_\text{n}}$ is the limiting velocity determined by the height 
$V_0$ of the optical potential barrier. This formula neglects the microscopic structure and roughness of the wall, which are typically included phenomenologically in the parameters 
$\eta$ (absorption coefficient) and $V_0$, allowing approximate applicability even for rough surfaces. No better approximation for UCN absorption by trap walls exists to date.

Most UCNs reflect elastically from walls, either specularly or diffusely. The probability of diffuse scattering is generally assumed to be 
$p_\text{d} \sim 0.1$. In Monte Carlo simulations, Lambert's cosine law is commonly used to model the angular intensity distribution of diffusely scattered UCNs \cite{Fomin2023,Fomin2019,Ayres2018,Fomin2018,Fomin2017,Serebrov2013MC}:
\begin{equation}
	I(\theta ) =I_0 \cos\theta ,  \label{Lambert}
\end{equation}
where 
$\theta$ is the angle between the direction of the scattered neutron velocity and the surface normal. However, Lambert's law is not universal, even in optics, with significant deviations observed \cite{LambertReview2020,Hecht1976,Mamouei2021}. For UCNs, no rigorous derivation of Lambert's law exists. Small-scale defects on trap walls, with characteristic sizes less than the UCN wavelength ($\lambda \sim 100$ nm), produce an angularly uniform distribution of diffuse scattering probabilities, rather than following Lambert’s law. Such defects include surface roughness, nanopores, impurities, etc. Even ultrasmooth surfaces, such as nickel or beryllium prepared for ultraviolet mirrors on satellites, exhibit submicron-scale roughness ($\lesssim 100$ nm) visible in atomic force microscopy or scanning electron microscopy \cite{Chkhalo2019}. Beryllium, previously used in UCN traps \cite{Ignatovich/1990,Ignatovich1996}, has now been replaced by hydrogen-free perfluoropolyether coatings such as Fomblin \cite{Serebrov2008PhysRevC.78.035505,Serebrov2018PhysRevC.97.055503}, which also exhibit strong submicron roughness \cite{Russell2007,Ganesh2014,Masciullo2018}.
Large roughness or pores with dimensions 
$d\gg \lambda $ scatter with small wavevector transfer 
$\sim 1/d$, broadening the specular reflection peak in the angular distribution of UCN velocities and deviating from Lambert’s law. Lambert’s law is supported by the detailed balance principle, which requires scattering probabilities proportional to 
$\cos\theta $, as the collision rate is also proportional to 
$\propto \cos\theta$ (see p. 96 of Ref. \cite{Ignatovich/1990}). Ignoring gravity, Lambert’s law indeed maintains isotropy in velocity distribution, tending toward isotropic redistribution of neutron velocities.

Lambert's law provides an isotropic velocity distribution only if the gravity force is negligible. This condition is correct, for example, for small traps closed at the top. For large traps, especially with an open top, our calculations show a strong deviation from an isotropic and uniform neutron velocity distribution even with Lambertian diffuse scattering of UCN from the material walls.
In large material UCN traps with complex curved walls used in neutron lifetime experiments \cite{Serebrov2008PhysRevC.78.035505,Serebrov2018PhysRevC.97.055503}, isotropy in neutron velocities is partly achieved through specular reflections from surfaces of varying orientation, independent of diffuse scattering. Consequently, measuring angular velocity distributions in such traps would provide limited insight into deviations from Lambert's law. However, theoretical calculations of angular velocity distributions after diffuse scattering on realistic trap walls are necessary to refine neutron absorption probabilities and corresponding limits on neutron $\beta$-decay lifetime precision 
$\tau_\text{n}$, which depends on the absorption angle as shown in Eq. (\ref{mu}).

For large material traps of simple geometries, such as rectangular or cylindrical shapes, our Monte Carlo simulations reveal that the UCN wall absorption rate depends on the diffuse reflection law due to its influence (even if it's weak) on angular velocity distributions. For realistic traps with complex geometries, this dependence may be less significant, as isotropy is enforced by reflections from variously oriented walls. Nevertheless, understanding deviations from Lambert’s law remains a fundamental question that requires experimental validation for different materials and geometries.

In this study, we propose an experimental setup to test Lambert’s law for UCN diffuse scattering on trap walls and quantify deviations. Detailed analysis and Monte Carlo simulations follow.

\textbf{2. Description and qualitative justification of the proposed experiment.}

\begin{figure}[tb]
	\centering
	\includegraphics[width=\linewidth]{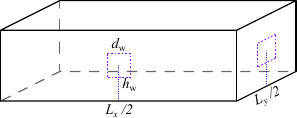}
	\caption{Fig. 1. (Color online) Elongated material trap for UCNs with dimensions $L_x \times L_y \times L_z$. Square windows (front and terminal
		) 
		of size $d_\text{w}$ at height $h_\text{w}$ are marked, through which neutrons escape from the trap.}
	\label{trap}
\end{figure}

The scheme of the proposed experiment to measure deviations from Lambert's law for neutrons is depicted in Fig.~\ref{trap}. UCNs are placed in an elongated material trap of size $L_x \times L_y \times L_z$, where $L_x \gg L_y , L_z$. The trap can also be cylindrical, resembling a tube typically used to transport neutrons from a source to an experimental setup. However, the Monte Carlo calculations below, which quantitatively describe the proposed effect, were performed for a rectangular trap. Unlike modern experiments to measure neutron lifetime~\cite{Serebrov2008PhysRevC.78.035505,Serebrov2018PhysRevC.97.055503}, the UCN trap dimensions are better chosen to be significantly smaller than the maximum neutron height $h_{\max}=V_0/m_\text{n} g$ in the gravitational field, and the top should not be open. Two small windows of equal area, front and terminal (see Fig.~\ref{trap}), are made in the trap. Detectors that count the neutrons escaping through these windows or tubes leading to such detectors are placed behind them. For successful measurement of deviations from Lambert's law, only the ratio of neutrons escaping through different windows, rather than their absolute numbers, is required. This simplifies the experiment and increases its precision.

If Lambert's law of diffuse neutron scattering in the trap holds, then for $L_z\ll h_{\max}$, where gravity can be neglected, the neutron velocity distribution will be isotropic and uniform (see page~96 in the book~\cite{Ignatovich/1990}), i.e., independent of position. Consequently, the number of neutrons escaping through the terminal and front windows will be identical. If Lambert's law is violated, the neutron velocity distribution will no longer be isotropic and uniform, and the numbers of neutrons escaping through different windows will differ.

Consider the case where, instead of Lambert's law (\ref{Lambert}), the diffuse scattering is isotropic and does not depend on the neutron emission angle: $I(\theta ) = \operatorname{const}$. Such isotropic scattering occurs if the scatterer size is much smaller than the wavelength of the incoming particle~\cite{LL3}, which for UCNs is quite large, $\gtrsim 100$ nm. In this case, since the number of neutrons incident on the wall is proportional to the cosine of the angle of incidence, $n_\text{in}\propto \cos\theta_\text{in}$, and the number of neutrons emitted upon isotropic diffuse scattering is independent of the velocity direction, more neutrons will travel along the trap walls. Due to the trap's elongated geometry, neutrons reflect much more frequently off its long walls, i.e., the front, rear, top, or bottom walls, than off its terminal walls. Consequently, under isotropic diffuse scattering, the number of neutrons $N_\text{e}$ escaping through the terminal window increases, while the number $N_\text{f}$ passing through the front central window decreases. We propose determining the ratio $N_\text{e}/N_\text{f}$ experimentally. If the measured ratio $N_\text{e}/N_\text{f}>1$, it indicates that Lambert's law is violated toward isotropic diffuse scattering. If $N_\text{e}/N_\text{f}<1$, it indicates a violation of Lambert's law in the opposite direction, with scattering occurring more along the normal to the surface.
For a quantitative evaluation of the proposed effect of neutron escape number differences through different trap windows, depending on the scattering law and trap geometry, we performed Monte Carlo calculations described below.

\textbf{3. Monte Carlo calculation.}
The material UCN trap modeled in our calculations is a long rectangular box with dimensions 
$L_y = L_z = 1$ cm, and its length $L_x$ varies up to 10 m (see Fig.~\ref{trap}). The box has square windows of size $d_\text{w} = 1$ mm at a height $h_\text{w} = 0.5$ cm.
The following parameters, close to real experiments~\cite{Serebrov2008PhysRevC.78.035505,Serebrov2018PhysRevC.97.055503} and other calculations~\cite{Fomin2019,Fomin2018}, were used: loss coefficient $\eta = 10^{-5}$, 
$v_{\lim} = 4.85$ m/s, diffuse scattering probability $p_\text{d} = 0.1$.
At the start of the Monte Carlo simulation, a neutron is located at a random position at the bottom of the trap. Its speed does not exceed $v_{\max} = 0.9 ~ v_{\lim}$ and has a random initial direction. The neutron moves within the trap according to the equations of motion in a gravitational field until it collides with the trap's ceiling/walls/bottom. Neutron hits on small windows (see Fig.~\ref{trap}) are counted.
A neutron can be absorbed by the wall with a probability $\mu (v_{\perp})$, given by equation~(\ref{mu}).
With probability $1-p_\text{d}$, the neutron reflects specularly; otherwise, it scatters diffusely. In this case, it may reflect according to Lambert's law with probability $p_\text{L}$ or scatter with an isotropic angular distribution with probability $1 - p_\text{L}$\footnote{The total probability of Lambertian scattering, provided the neutron is not absorbed, is evidently $p_\text{d}p_\text{L}$, and that of isotropic scattering is $p_\text{d}(1-p_\text{L})$.}.

\begin{figure}[tb]
	\centering
	\includegraphics[width=\linewidth]{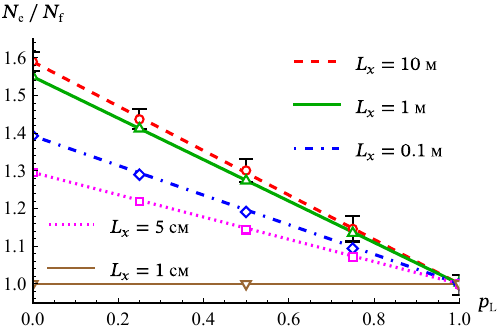}
	\caption{Fig. 2. (Color online) Ratio of the number of neutrons passing through the terminal 
	window $N_\text{e}$ to the number of neutrons passing through the front window $N_\text{f}$ as a function of the Lambert scattering probability $p_\text{L}$, for traps of various lengths $L_x$.
	For $L_x = 10$~m, the standard deviation is shown, averaged over 10 simulations of $10^4$ neutrons.}
	\label{pL_fig}
\end{figure}
By calculating the ratio of the number of neutrons $N_\text{e}$ counted in the terminal window to the number $N_\text{f}$ counted in the front window for a large number of neutrons ($N \gtrsim 10^4$) at various $L_x$ and $p_\text{L}$, we obtained the results shown in Fig.~\ref{pL_fig}. These results confirm our qualitative arguments that the probabilities of neutron escape through the terminal and front windows are equal only for Lambertian diffuse scattering, i.e., only when $p_\text{L}=1$. Even for small deviations from Lambert's law, the difference between $N_\text{e}$ and $N_\text{f}$ is noticeable and measurable. This difference amounts to tens of percent and grows linearly with the deviation, described by the parameter $1-p_\text{L}$. The ratio $N_\text{e}/N_\text{f}$ increases with the trap's elongation but remains significant even if the linear dimensions ratio of the trap is $L_x/L_y =5$. Further lengthening the trap while keeping its width and height constant has less impact. For instance, for $L_x/L_y =1000$, the difference $N_\text{e}-N_\text{f}$ is only twice that for $L_x/L_y =5$.

\begin{figure}[tb]
    \centering
    \includegraphics[width=\linewidth]{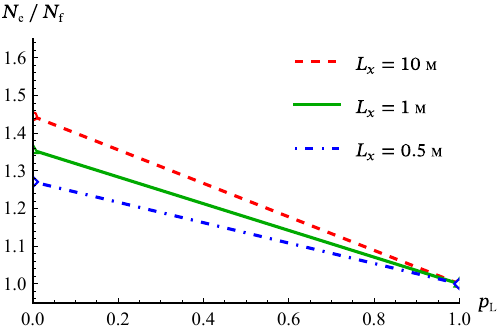}
    \caption{Fig. 3. (Color online) The same as Fig. \ref{pL_fig} for $L_y=L_z=10$ cm.}
    \label{pL_fig2}
\end{figure}

To ensure that the obtained result weakly depends on the trap size, provided the trap height remains below the maximum height of neutrons $h_{\max}=v_{\max}^2/2g\sim 1$ m determined by their energy and gravity, we also performed calculations for larger trap dimensions $L_y = L_z = 10$ cm. The results, shown in Fig. \ref{pL_fig2}, indicate only a slight weakening of the effect and demonstrate that the ratio $N_\text{e}/N_\text{f}$ continues to increase linearly with the deviation $1 - p_\text{L}$ from Lambert's law, being geometrically determined by the ratio $L_x/L_y$ rather than the trap size.

\textbf{4. Discussion.} We have proposed a method for studying the angular distribution of diffuse reflection of ultracold neutrons from material walls. Quantitative estimates of the proposed effect—the difference in the number of neutrons exiting through different openings of the UCN trap depending on their diffuse scattering law and trap geometry—obtained from Monte Carlo simulations, show a significant and easily measurable effect, amounting to tens of percent. If the measured difference in the number of neutrons exiting through different openings of a long trap is small, such an experiment is also an important physical result, as it confirms Lambert's law for diffuse scattering of ultracold neutrons.

The proposed experiment does not require a high neutron concentration as in neutron lifetime measurements \cite{Serebrov2008PhysRevC.78.035505,Serebrov2018PhysRevC.97.055503,Ezhov2018,Gonzalez2021}, since it does not involve keeping neutrons in the trap for a long time or determining the exponent of the time dependence of their concentration. It is only necessary to find the ratio of the number of neutrons exiting through the end and side openings of equal area. To eliminate possible errors associated with different opening sizes and detector locations, the measurement results can first be calibrated using a cubic trap with $L_x = L_y = L_z$, for example, by placing shutters. In this case, the number of neutrons exiting from different openings should be the same, regardless of the diffuse scattering law (see the solid brown line in Fig. \ref{pL_fig}).

As shown in Figs. \ref{pL_fig} and \ref{pL_fig2}, the proposed effect weakly depends on the trap size and shape, provided that (1) the trap length is significantly greater than its transverse dimensions and (2) the trap walls remain straight along the $x$-direction. However, the magnitude of the proposed effect may be lower than our theoretical predictions if the internal trap surfaces have significant large-scale roughness $\gtrsim 1$ $\textmu$m, which, due to multiple specular reflections, mix different components of the neutron velocity. Therefore, for a successful experiment, it is desirable to make the trap walls as straight as possible along its length.

The proposed measurement of deviations from Lambert's law in the diffuse scattering of UCN from material walls could be crucial for improving the accuracy of estimating UCN losses in material traps, and thus for precise measurements of the $\beta$-decay rate or neutron lifetime $\tau_\text{n}$. The main alternative to $\tau_\text{n}$ measurements in material traps are magnetic traps with an open top, which currently have nominally higher precision \cite{Gonzalez2021}. Nevertheless, magnetic traps also face several challenges \cite{Ezhov2018,GONZALEZALONSO2019165,Ezhov2023}, such as accurately estimating the depolarization (spin-flip) rate in an inhomogeneous magnetic field and associated UCN losses. It is possible that the actual systematic uncertainty in $\tau_\text{n}$ measurements in magnetic traps is higher than the $\sim 0.2$ seconds estimated in \cite{Gonzalez2021}.

The study of diffuse scattering of UCN from material walls is interesting in itself, as there are no strict grounds for the applicability of Lambert's law to neutrons, and measuring deviations from it could provide deeper insight into the microscopic processes occurring during neutron-wall interactions. This fundamental question has so far been little studied, and we propose a fairly simple and useful experiment for such research.

\medskip

This work was supported by the RSF grant No. 23-22-00312.

\bibliographystyle{apsrev4-2}
%

\end{document}